\documentclass[review]{elsarticle}
\usepackage{graphicx} \usepackage{amsmath}
\usepackage{lineno,hyperref}
\modulolinenumbers[5]

\journal{Communications in Computational Physics }

\begin{document}

\begin{frontmatter}

\title{Numerical path integral approach to quantum dynamics and stationary quantum states}

\author{Ilkka Ruokosenm\"aki and Tapio T.~Rantala}
\address{Department of Physics, Tampere University of Technology, Finland}
\ead{FirstName.LastName@tut.fi}

\begin{abstract}
Applicability of Feynman path integral approach to numerical simulations of
quantum dynamics in real time domain is examined. Coherent quantum dynamics is demonstrated with one dimensional test cases (quantum dot models) and performance of the Trotter kernel as compared with the exact kernels is tested.  Also, a novel approach for finding the ground state and other stationary sates is presented.  This is based on the incoherent propagation in real time.  For both approaches the Monte Carlo grid and sampling are tested and compared with regular grids and sampling.  We asses the numerical prerequisites for all of the above.
\end{abstract}


\end{frontmatter}


\section{Introduction}

Feynman path integral (PI) approach offers an intuitively welcome description of nonrelativistic quantum mechanics \cite{feynman,feynman2}, where classical mechanics emerges transparently from disappearing wave nature of particles along with vanishing Planck constant.  
In PI approach the presentation of the quantum dynamics with a propagator also in stationary quantum states is transparent, in contrast with the conventional approaches, where time evolution is seen in the phase factor, only.  However, working out analytical or  computational solutions to practical problems becomes more demanding with PI \cite{kleinert,schulman}, and obviously, this is one of the main reasons for path integrals not being a popular choice for considering quantum dynamics, not to mention the stationary quantum states.  

For the above reasons the dynamical phenomena in nonrelativistic quantum mechanics are conventionally considered by searching or simulating solutions to the time dependent Schr\"odinger equation.  This is almost trivial for a single particle, but becomes laborious and needs a number of approximations with growing complexity in a many-body system.  In contrast, with PI the many-body interactions are included transparently and exactly within numerical accuracy.  Nevertheless, the PI approach is rarely used outside quantum field theory or without Monte Carlo (MC) technique as the working horse.

However, it is worth mentioning that PIMC has proven to be very successful in simulations of periodic imaginary time propagation of many-particle systems, which leads to the finite temperature equilibrium statistical physics  description of the many-particle system in terms of mixed state density matrix \cite{ceperley, kylanpaa}.  By treating all particles with the same PIMC approach it is possible to evaluate the finite temperature electronic structure with exact account of many-body effects and beyond Born--Oppenheimer approximation as demonstrated, already \cite{kylanpaaP, militzer}.
PIMC is also robust enough to be used in various applications in nanoscience \cite{weiss, gull}.

Beyond the analytical solutions to stationary states or quantum dynamics, which are very few \cite{kleinert,schulman,makri1, makri2}, numerical simulation of coherent real time propagation faces substantial challenges related to the interference of paths: how to choose or sample the relevant paths in a balanced way, i.e.~weighting the ones with most contribution through constructive interference and avoiding waste of efforts to those with negligible contribution due to destructive interference.  In practice, time evolution of the complex many-body wave function in a space with high number of dimensions leads to even higher dimensional path integrals, which obviously can be sampled efficiently with the Monte Carlo technique, only.  There, the interference related slow convergence has been called as "numerical sign problem" \cite{makri1, makri2} or phase (sign) problem.
Sophisticated "stationary phase weighting" methods have been developed to overcome this without Monte Carlo technique \cite{filinov, wang}.

There are still no preferable solutions to these problems, although many approaches and approximations for certain types of systems have been found \cite{makri3,marchioro}.  Basically these methods rely on effective propagators \cite{makri4} with desired properties.  They are relatively well behaving and use the advantageous features of the PI formalism, e.g., reduction of the total system into two parts: the lower dimensional system of interest and the effect of an environment modeled with an influence functional \cite{feynman}.  Often, the effect of the environment can be approximated classically, leaving only a lower dimensional system to be inspected quantum mechanically.  Such methods have been shown to be successful in evaluation of the time evolution of a quantum--classical many-body systems \cite {makri5} for heavier particles than electrons.

Since there is no perfect method for solving dynamical full quantum many-body problems in practice, it is useful to look at different methods, how they can be used, what are their strengths and weaknesses and what is needed in implementation of those methods.

In this paper, we deal with real time quantum dynamics with both coherent and incoherent propagation.  Next, we present the basic theory and the approximative Trotter kernel, and in sec.~3, the numerical approach to evaluation of propagation and expectation values.  In sec.~4 we define one dimensional electron-in-quantum-dot models chosen for testing.  In sec.~5 we analyze results for coherent quantum dynamics and in sec.~6 we finally present a novel approach to search for stationary quantum states and the ground state, in particular.  The last section presents our conclusions.

\section{Path integral and propagators}

Consider non-relativistic particle propagation in one, two or three dimensional space \(\Omega\) from \(x_a\) to \( x_b\) in time interval from \(t_a\) to \( t_b\) along all possible paths \(x(t)\).  The path integral over all paths defines the propagator
\begin{equation}    \label{kernel}  
	K(b,a) = \int_{a}^{b} \exp\left[ \frac{\rm i}{\hbar} S_x[b,a] \right] \mathcal{D}x(t),
\end{equation} 
where \(S_x[b,a] = \int_a^b L_x {\rm d}t  \) is the action of the path \(x(t)\) from \(a=(x_a,t_a)\) to \(b=(x_b,t_b)\) and \( L_x \) is the corresponding Lagrangian \cite{feynman,feynman2}.
Time evolution of the probability amplitude, i.e., the wave function \(\psi (x, t)\) in space \(\Omega\) can now be written as
\begin{equation}   \label{propag}  
	\psi (x_b, t_b) = \int_{\Omega} K(x_b,t_b;x_a,t_a) \psi (x_a, t_a) {\rm d} x_a,
\end{equation}
where \(t_a < t_b\).
From this relation the time dependent Schr\"odinger equation can be derived \cite{feynman}, or alternatively, the time dependent wave function \(\psi (x, t)\) can be directly evaluated from the initial state \(\psi (x_a, t_a)\), in case the kernel \(K(x,t;x_a,t_a)\) is known.

However, general explicit forms of the propagator are known for simple cases, only, such as the particle with mass \(m\) in the one dimensional constant linear potential \( V(x) = - fx \),
\begin{eqnarray}  \label{linpropag}   
K(x_b,x_a; t) = \left[\frac{m}{2 \pi {\rm i} \hbar t}\right]^{1/2} 
	\exp \left[\frac{{\rm i}}{\hbar}(\frac{m}{2 t} ( x_b - x_a )^2 - \frac{t}{2}(V(x_a) + V(x_b)) - \frac{t^3 f^2}{24m} \right],
\end{eqnarray}
which reduces to the free particle propagator with \( f = 0 \) \cite{feynman}.
For the one dimensional forced harmonic oscillator
\begin{equation}  \label{harmpot}   
	V(x,t) =  \frac{m \omega^2}{2} x^2 - f(t) x
\end{equation}
the exact explicit propagator takes the form \cite{feynman}
\begin{equation}  \label{harmfpropag}   
	K(x_b,x_a; t) = \left[\frac{m \omega}{2 \pi {\rm i} \hbar \sin(\omega t)}\right]^{1/2} 
	\exp \left[ \frac{{\rm i}}{\hbar} S_{\rm cl}  \right],
\end{equation}
where \( S_{\rm cl} \) is the classical action.  For \( f \equiv 0 \) this is
\begin{equation}  \label{harmpropag}   
	S_{\rm cl} = \frac{m \omega}{2 \sin(\omega t)} \left[ (x_b^2 + x_a^2) \cos(\omega t) - 2 x_b x_a \right].
\end{equation}

For numerical approaches robust approximations are needed.  It is advantageous that also in nontrivial forms of potential the propagation is straightforward to evaluate and with increasing numerical accuracy the propagator approaches the exact limit.  With this in mind we discretize the time \( t = t_b - t_a \) to a number of short steps \(\Delta t \).
This is straightforward, because
\begin{equation}   
	K(b,a) = \int_{\Omega} K(b,c)K(c,a) {\rm d}x_c,
\end{equation}
for  \(t_a < t_c < t_b\).   This follows from additivity of action \( S[b,a] = S[b,c] + S[c,a] \) for any path \cite{feynman}.

Now, with a small \(\Delta t \) the quantum paths can be expected to give the main contribution close to the classical path, for which \(\Delta x = x_b - x_a \) is also small.  This follows from the canceling kinetic energy \( T \) contributions due to the destructive interference of paths in long path propagation.  This presumes, of course, smooth enough potential \( V \), for which also the commutator \( [T,V] \) is small.

Furthermore, for numerical approaches it is essential that the chosen discretization also converges to the exact formalism at the limit \( \Delta t \rightarrow 0 \), and the faster the better for practical purposes.  Also, it is preferable that computational efforts are not wasted for computation of almost canceling contributions more than needed for the chosen target accuracy.

Now, Eq.~(\ref{linpropag}) gives numerically useful approximation, which can be further simplified by neglecting the last term, cubic in \( \Delta t \), for short enough time steps.  Thus, we arrive at the symmetrized Trotter kernel \cite{makri1, makri2}
\begin{eqnarray}  \label{trottpropag}  
   K(x_b,x_a; \Delta t) \approx \left[\frac{m}{2 \pi {\rm i} \hbar \Delta t}\right]^{D/2} \exp \left[\frac{{\rm i}}{\hbar}(\frac{m}{2 \Delta t} ( x_b - x_a )^2 - \frac{\Delta t}{2}(V(x_a) + V(x_b)) \right] ,
\end{eqnarray}
where \( D \) is the dimensionality of space.  

This propagator can also be found from the hamiltonian formulation \cite{schulman}.
For a time independent hamiltonian \(H = T + V\), where \(T\) and \(V\) are the kinetic and potential energies, the propagator can be written as \cite{schulman}
\begin{equation}  \label{TVpropag}  
K(x_b,x_a;\Delta t) = \langle x_b | \exp[-\frac{{\rm i}}{\hbar}H \Delta t ] | x_a \rangle = \langle x_b | \exp[-\frac{{\rm i}}{\hbar}(T + V)\Delta t]| x_a \rangle,
\end{equation}
where \( \Delta t = t_b - t_a \).  
Now, by using the Zassenhaus formula \cite{schulman,suzuki}
\begin{eqnarray}   
\begin{split}
\exp[-\frac{{\rm i}}{\hbar}(T + V) \Delta t ] = & \exp\left[-\frac{{\rm i} \Delta t}{\hbar}T \right]
 \exp\left[-\frac{{\rm i} \Delta t}{\hbar}V \right] \times \\
 & \times \exp\Big\{ \left(\frac{{\rm i} \Delta t}{\hbar}\right)^2 \frac{[T,V]}{2} \Big\} 
 O\Big\{ 1 + \left(\frac{{\rm i} \Delta t}{\hbar}\right)^3 \Big\}
 \end{split}
\end{eqnarray}
and by neglecting factors which approach one in the second order or higher in \( \Delta t \), as \( \Delta t \rightarrow 0 \), and using the path integral formulation, we arrive at the approximation (\ref{trottpropag}). Thus, this approximation is accurate almost to the second order in \( \Delta t \) for a smooth potential with \( [T,V] \rightarrow 0\) as \( \Delta x \rightarrow 0\) or  \( \Delta t \rightarrow 0\).  In fact, this is what the kernel in Eq.~(\ref{linpropag}) also suggests.

Clearly, in numerical approaches it is the kinetic energy part, which brings in the challenges as \( \Delta t \rightarrow 0\), but as pointed out above, already, the resulting large momentum -- short wave length oscillations of the propagator interfere destructively and should be damped out without wasting computational efforts.  The potential energy part behaves the opposite way with respect to the time step, and becomes laborious only in case of large potential gradient at possible singularities in the potential function.

We consider and test the Trotter kernel Eq.~(\ref{trottpropag}) against the exact kernels Eqs.~(\ref{linpropag}) and (\ref{harmfpropag}) in numerical simulations of one-dimensional harmonic oscillator (ODHO) and quantum well (QW), both in stationary eigenstates and wave packet propagation.

\section{Numerical evaluation of propagation and expectation values}

Numerical evaluation of the integral Eq.~(\ref{propag}) is the core problem, here.  For that, we span grids  \( {\bf g}_a = \{{x_a}_i\}_{i= 1}^{N_a} \) and  \( {\bf g}_b = \{{x_b}_j\}_{j= 1}^{N_b} \) for wave functions at \(a\) and \(b\).  It is practical to define the grid density profiles or distribution functions \( g_a(x) \) and \( g_b(x) \), as (possibly normalized) inverse average grid spacing.  With small enough time step \( \Delta t \) we can assume the same restricted space \( \Omega \) for both \( \psi_a \) and \( \psi_b \), and for simple cases, also the same grid \( {\bf g} = {\bf g}_a = {\bf g}_b \) with the same size \( N = {N_a} = {N_b} \).

The simplest equally spaced regular grid, i.e., with \( g \) constant, between end points may generate fake constructive diffraction patterns.  This is the diffraction grating effect, which can be removed out by increasing the grid size \( N \). 
Usually, a better choice is some other regular distribution of \( g \), like gaussian or some other, related to the probability density or (the absolute value of) the wave function, itself.

Of course, Monte Carlo grids with given distributions \( g \) serve well, if smooth and sizable enough.  There are methods for the analysis of "smoothness" of the distribution, such as Kolmogorov--Smirnov test \cite{ks}.  In fact, with the increasing number of dimensions Monte Carlo grids may remain as the only practical choice.
Further smoothing and averaging out accumulative errors is attained with a continuous random change of the MC grids, within the predefined density profiles.  For restricted range of dynamics, it may be practical to use identical distributions, i.e., \( g_a(x) = g_b(x) \), but  \( {\bf g}_a \ne {\bf g}_b \).

Ongoing random evolution of \( \{x_i\}_{i= 1}^{N_i} \) also means sampling of continuous space, instead of a discrete grid.  This evolution can be adapted to follow the time evolution of the wave function or some related distributions like the absolute value or the probability distribution of the wave function, i.e., \( g(x,t) \propto | \psi(x,t) |^n \), \( n = 1 \) or \( 2\), for example.

The distribution function \( g(x) \) appears as an inbuilt weight factor in the integration of Eq.~(\ref{propag}).  In the one-dimensional space it is straightforward to write \( g(x) = {\rm d} G(x)/ {\rm d} x \), in terms of the cumulative distribution function \( G \).  Thus,  Eq.~(\ref{propag}) becomes in form \( \psi(b) = \int_0^1 K(b,a) \, \psi(a) \ g_a^{-1}(a) \, {\rm d} G_a \).  For  propagation over the time interval \( \Delta t = t_b - t_a \) with \( t_a = 0 \), numerical calculation can be carried out as
\begin{eqnarray}   \label{wfpropag}  
\begin{split}
 \psi (x_j, \Delta t) 
 &= \int_0^1 K(x_j,\Delta t;x_i,0) \frac{\psi (x_i, 0)}{\ g_a(x_i)} {\rm d}G_a(x_i) \\
 & \approx \sum_{i=1}^{N_a} \frac{K(x_j,x_i;\Delta t) \psi (x_i, 0)}{g_a(x_i)}.
 \end{split}
\end{eqnarray}

Hence, it seems obvious that \( \psi(a) \) should decay faster than \( g_a \) in order to avoid numerical instabilities.  For real \( \psi(a) \) or for its absolute value this can be easily established, whereas for the two parts of complex \( \psi(a) \) this can be expected to be more tricky.  The phase factor of calculated \( \psi(b) \) relates to the "local total energy", and therefore, it serves as a good indicator of numerical stability.  Therefore, it seems possible to find phase factor based algorithms for stabilization of propagation and for removing numerical errors. 

In principle, the distribution \( g_a(x) \) needs not to be known analytically, if \( g_a(x_i) \) can be evaluated from the wave function, for example.  Furthermore, negative sign can be assigned to \( g_a(x) \) at some range of \( x \), if relevant for some reason.

Monte Carlo evaluation of expectation values of local operators, like the multiplicative potential \( V(x) \), at time \( t_a \), can be done with
\begin{equation}   \label{ExptV}  
	\langle V \rangle = \int_0^1 \frac{\psi^\star(x_i, t) V(x_i) \psi(x_i, t)}{\ g(x_i)} {\rm d} G(x_i) \approx \sum_{i=1}^{N} \frac{V(x_i) | \psi (x_i, t) |^2}{g(x_i)},
\end{equation}
where the operator can be time dependent, too.

Similarly, we calculate the total energy from
\begin{equation}   \label{ExptE}  
	\langle E \rangle \approx \sum_{i=1}^{N} \frac{E_L(x_i) | \psi (x_i, t) |^2}{g(x_i)},
\end{equation}
where the local energy is evaluated from the increase in wave function phase \( - \Delta \phi(x) \) within a time step \( \Delta t \) as \( E_L(x) = - \Delta \phi(x) \hbar / \Delta t \).
Then, the kinetic energy \( \langle T \rangle \) can be evaluated from \( \langle E \rangle = \langle T \rangle + \langle V \rangle \).

\section{One-dimensional harmonic oscillator and quantum well}

We first consider the one-dimensional harmonic oscillator (ODHO), i.e., a particle in the potential of Eq.~(\ref{harmpot}) with \( f(t) \equiv 0 \).  Thus, 
we have the time-independent potential
\begin{equation}  \label{harmopot}   
	V(x) = \frac{1}{2} m \omega^2 x^2.
\end{equation}
We choose the parameters describing an electron in an atom size "quantum dot" to maximize the quantum effects and challenge for simulation of dynamics.  We use atomic units, where  \(\hbar = 4\pi \varepsilon_0 = e = m = a_0 = 1\), the last three being the charge, mass and Bohr radius of the electron.  This leads to the atomic unit energy of Hartree, Ha \( = \hbar^2 / (m a_0^2) \approx 27.211384 \) eV, which also defines the unit of the potential in Eq.~(\ref{harmopot}).  The atomic time unit becomes as  \( t_0 = (m a_0^2) / \hbar \approx 24.18884 \times 10^{-18}\) s \( \approx 24\) as.

Now, by substituting \( m = 1 \) and \( \omega = 0.1 \) (\( = \hbar \omega \)), we have the corresponding eigenenergies \( E_\nu \) with equal contributions from kinetic and potential energies and eigenstates \( \psi_\nu(x) = (2^\nu \, \nu! / \sigma_0)^{-1/2} \pi^{-1/4} H_\nu(x/\sigma_0) \exp(-x^2 / 2\sigma_0^2)  \), where \( H_\nu \) are Hermite polynomials and \( \sigma_0 = \sqrt{\hbar / m \omega} \approx 3.16 \).  For the ground state we have \( \psi_0(x) = \pi^{-1/4} \sigma_0^{-1/2} \exp(-x^2 / 2\sigma_0^2) \) and \( E_0  = 0.050 \).  Thus, \( E_1  = 0.150 \).

The one-dimensional quantum well (QW) or "particle in a box" 
\begin{eqnarray}  \label{ODPBpot}   
  V(x) = 
 \begin{cases} 
   0 & \mbox{for}   |x| < L/2   \\
   \infty & \mbox{otherwise,} 
 \end{cases} \mbox{   and}
\end{eqnarray}
with \( L = 20 \) is also used as a test case, where relevant.  Here, we have the free particle eigenstates with energies \( E_\nu = \frac{1}{2}k^2 \), where \( k = 2\pi / \lambda \) and \( \nu \lambda / 2 = L \).  Thus, \( E_1 = \frac{1}{2} (\pi / L)^2 \approx 0.01234 \) and \( E_2 = 2 (\pi / L)^2 \approx 0.04935 \).

\section{Coherent dynamics}

\subsection{Stationary states}

First, we searched for numerical parameters, which keep the eigenstates stationary with an acceptable accuracy.  The three lowest eigenstates of ODHO (\( \hbar \omega = 0.1 \)), Eq.~(\ref{harmopot}), turn out to remain stable in a simulation with an even spaced grid of size \( N = 10^3 \) in the domain \( -12 < x < 12 \) with the time step \( \Delta t = 1 \).  The potential energy expectation value (\ref{ExptV}) fluctuates around the time average \(\overline{\langle V_0 \rangle} = 0.02503 \) with a standard deviation \( \sigma \approx 3 \times 10^{-5} \), and correspondingly, the total energy (\ref{ExptE}) becomes as \(\overline{ \langle E_0 \rangle } = 0.05002 \) with \( \sigma \approx 4 \times 10^{-9} \).  Thus, a small grid related error remains.

We find that the time step should be small enough (\( \Delta t_{\rm max} \approx 4 \)) to justify the Trotter approximation, Eq.~(\ref{trottpropag}), for ODHO.   Shortening the time step calls for more accurate grid due to increasing kinetic energy, i.e., oscillatory nature of the exponential in Eq.~(\ref{trottpropag}). The potential energy contribution to phase oscillations is roughly two orders of magnitude less.  In general, we found the maximum time step and even grid size proportion to be related roughly as \(\Delta t_{\rm max} \times N \geq 10^3 \) for the Trotter kernel, Eq.~(\ref{trottpropag}).

The exact kernel Eqs.~(\ref{harmfpropag}--\ref{harmpropag}) of ODHO, however, allows unlimited time step and the accuracy depends on the grid, only.  Even so, the time steps of a multiple of half oscillation period can not be used, because \( \sin(\omega t) \) in the denominator causes divergence of both (\ref{harmfpropag}) and (\ref{harmpropag}).  With other time steps \( 1 \le \Delta t \le 500 \) and  \( N = 10^3 \) the potential energy keeps correct in \( 5 \) digits.  The total energy \(\overline{ \langle E_0 \rangle } \) becomes evaluated with same accuracy.

For the QW with constant potential the Trotter kernel is nearly exact \cite{schulman}.  However, numerical accuracy suffers from inaccurate description of discontinuities of the potential function Eq.~(\ref{ODPBpot}) at \( |x| = L/2 \).  Thus, the accuracy is limited by the grid spacing \( \Delta x \).  Obviously for this reason, we found the time propagation to be somewhat unpredictable.

For this case, we found that the Monte Carlo grid with a constant distribution function to solves the problem.  Time evolution of the grid, with \( g(x) = {\rm constant} \), samples the space continuously.  We found the grid size \( N = 10^3 \) sufficient for a stable simulation of the ground state in a QW \( L = 20 \) with the total energy \(\overline{ \langle E_0 \rangle } \) accurate in a few digits, for a few steps, already.  Obviously, other non divergent but adapted distributions \( g(x) \) will perform even better.

\subsection{Wave packet propagation}

Next, we consider real time evolution of gaussian wave packet oscillation in the harmonic potential (ODHO), above.  As a test case we use the Glauber state, also called coherent or quasi-classical state, because of classical like oscillation retaining the wave packet shape rigid.  In fact, the width of the Glauber state gaussian is that of the ground state, in the present case \( \psi(x) = \pi^{-1/4} \sigma_0^{-1/2} \exp(-x^2 / 2\sigma_0^2)  \).  The oscillation frequency is, of course, \( \omega = 0.1 \) and period \( T = 2\pi/\omega \approx 62.83 \), for any oscillation amplitude \( A \).

With the Trotter kernel and grid size \( N = 10^4 \) the time step dependence is small.  With \( A = \sqrt{20} \) and starting from rest, the total energy is that of the first excited state, see Fig.~1. 
Both \( \Delta t  = 2 \pi/60\) and \( \Delta t  = 2\pi / 200 \), and wave packet propagation of one period leads to potential energy error of \( -0.0027 \), only.  With the exact kernel, Eqs.~(\ref{harmfpropag}--\ref{harmpropag}), arbitrarily long time steps can be taken, except those, for which \( \sin(\omega \Delta t) \approx 0 \), as pointed out above.

\begin{figure}[t]
\label{glauber}
\includegraphics[trim=2cm 0cm 2cm 2cm, clip=true,scale=0.45]{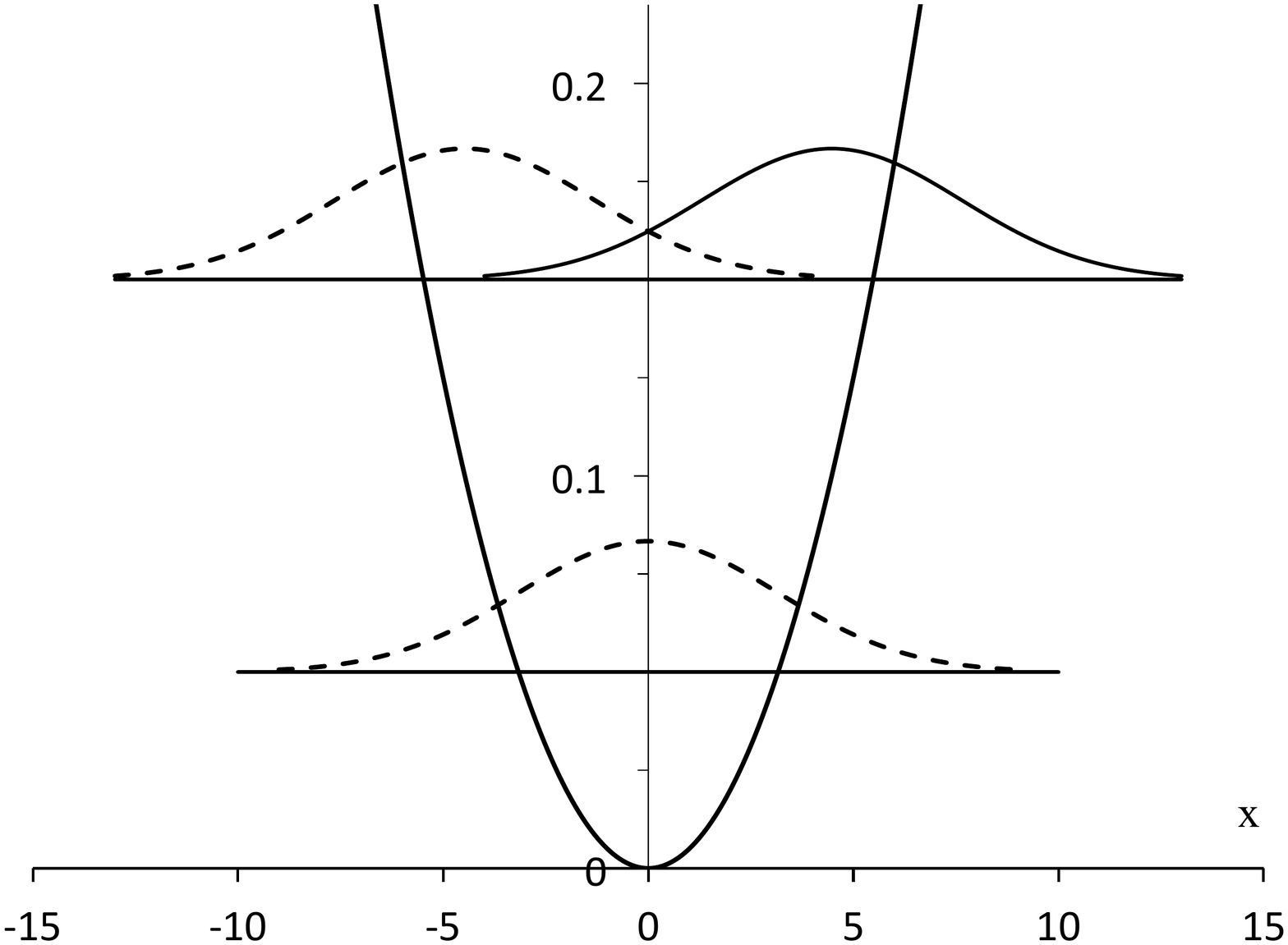}
\caption{The ODHO potential and the starting Glauber state (full curves).  Dashed curves show the two other extreme phases of oscillation.  Horizontal lines indicate the ground and the first excited state energies.}
\end{figure}

\section{Incoherent dynamics}

\subsection{Stationary state search}

With the path-integral approach, simulation of stationary eigenstates is no more trivial than that of explicitly time dependent wave functions.  In both cases full propagation in the whole space needs to be similarly considered within each time interval.  This points to the inherent nonlocality of the wave function and quantum phenomena, in general.

An arbitrary pure quantum state can be expanded as a superposition of stationary eigenstates as \( \Psi = \sum_{k} c_k \psi_k \) and its time evolution in \( \Delta t \) is \( \Delta \Psi = \sum_k \exp(-{\rm i} E_k \Delta t ) c_k \psi_k  =   \sum_k [\cos( E_k \Delta t) -{\rm i}\sin( E_k \Delta t)] c_k \psi_k  \). By using the small angle approximation for short enough \( \Delta t \), this can be written as \( \Delta \Psi \approx \sum_k [1 - ( E_k \Delta t )^2/2 -{\rm i}( E_k \Delta t)] c_k \psi_k   \).

Consider now stepwise decoherence of the wave function in each time step, that is driven by removal of the small imaginary part.  Such incoherent time evolution,
\begin{equation} \label{uncoherent}   
	\Delta \Psi(\Delta t) = \sum_k [1 - ( E_k \Delta t )^2/2)] c_k \psi_k ,
\end{equation}
converges to quantum Zeno propagation at the limit \( \Delta t \rightarrow 0 \), if the eigenstate is real.  However, with a finite but short enough \( \Delta t \) it increases the contribution of the eigenstate with smallest absolute eigenvalue with respect to the chosen reference energy, if \( E_k \Delta t << 1 \) for all \( k \).  At the end, this state dominates and contributions from the other states die out.

This is what we call incoherent propagation, here, and demonstrate the respective time evolution in ODHO with the Trotter propagator in evenly spaced grid, see Fig.~2. 
Incoherent evolution depends on the initial state as shown.  In case where the ground state \( \psi_0 \) contribution is initially considerable, \( c_0 \neq 0 \), the convergence is fast.  However, in case where initially \( c_0 = 0 \), lowest of the states contributing to the initial wave function is found.  The ground state is found only after a small seed of \( \psi_0 \) has been sown from numerical errors in propagation.

\subsection{Ground state evaluation}

Finally, we consider accurate evaluation of the ground state, or another stationary state, after first finding it by the "stationary state search" described in the previous section.  With the incoherent propagation in ODHO by using the Trotter propagator we found accuracy of about five digits for the ground state energetics, independent of the grid size (\( N = 10^3 \) to \( 3 \times 10^4 \)) and accidentally with the time step \( \Delta t \approx 0.3 \).  Obviously, there remains a systematic error due to the grid and propagator.

Therefore, we again employ the Monte Carlo grid to sample the continuous space.  We also simplify the propagation, Eq.~(\ref{wfpropag}), to increase accuracy in the spirit of diffusion Monte Carlo (DMC) approach, where it is the distribution of walkers, which is the target ground state wave function.  This allows comparison of our approach to DMC, which is known as a robust and accurate method for finding and evaluation of properties of the ground state.

\begin{figure}[t]
\label{search}
\includegraphics[trim=2cm 0cm 2cm 1cm, clip=true,scale=0.45]{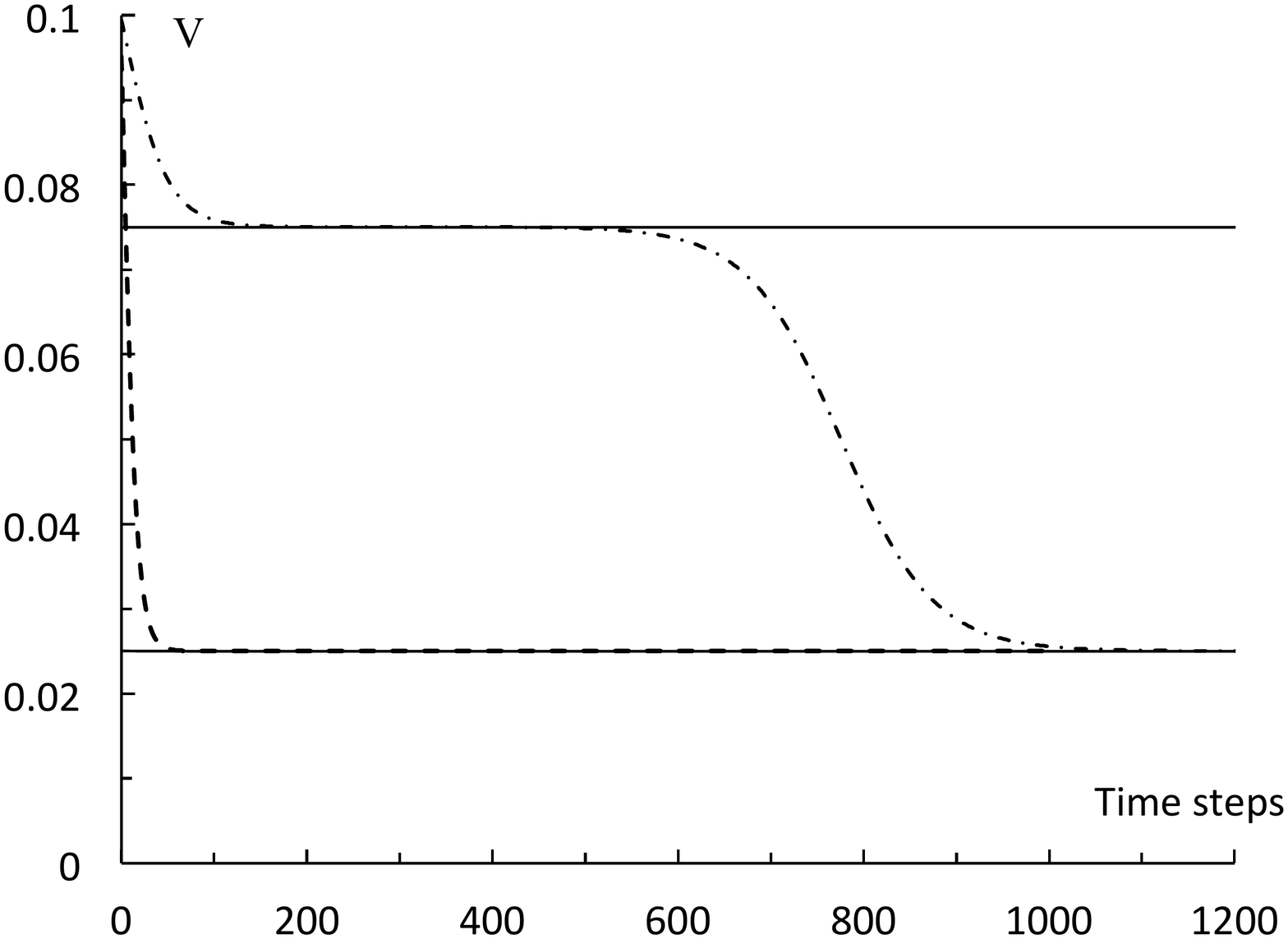}
\caption{Incoherent evolution of the superposition states to the ground state.  Dashed line starts from the superposition of the ground and 3rd excited state, whereas the dash dotted line starts from the superposition of the 1st and 2nd excited states.  Solid lines show the potential energies of the ground and 1st excited states.}
\end{figure}

Close enough the ground state we set \( g(x) = \psi (x) \approx \psi_0 (x) \), and consequently, approximate Eq.~(\ref{propag}) and (\ref{wfpropag}) for numerical Monte Carlo evaluation as
\begin{eqnarray}   \label{wghtpropa}  
\begin{split}
  \psi (x_j, \Delta t) 
  & = \int K(x_j,\Delta t;x_i,0) g(x_i) {\rm d} x_i \\ 
  & = \int_0^1 K(x_j,\Delta t;x_i,0) {\rm d} G(x_i) \approx \sum_{i=1}^{N_a} K(x_j,x_i;\Delta t),
  \end{split}
\end{eqnarray}
and therefore, \( \{x_i\}_{i= 1}^{N_a} \) are random numbers from distribution \( g(x) \) with the cumulative distribution function \( G(x) \), as discussed above.  Thus, in practice we run incoherent propagation
\begin{equation}   \label{wpropag}  
	\psi (x_b, \Delta t) = \int K(x_b,\Delta t;x_a,0) \psi (x_a, 0) {\rm d} x_a  
	,
\end{equation}
without an explicit starting amplitude \( \psi (x_a, 0) \), but hidden in the walker distribution, and
assuming good convergence of the distribution to the ground state wave function.  To sample continuous space, Metropolis Monte Carlo (MMC) can be used to carry out evolution of the walker distribution \( g(x) \), and if needed, stability can be increased by using the "time average" \( \overline{g(x)} \) from a longer simulation and partly overlapping grids \( {\bf g}_a = \{{x_a}_i\}_{i= 1}^{N_a} \) and \( {\bf g}_b = \{x_j\}_{j= 1}^{N_b} \), with \( N_a = N_b = N \).

\begin{table} [b]
\label{table}
 
\caption {Incoherent propagation in MC grid of the ODHO ground state with Trotter kernel. \( N \) is the grid size, \( \Delta t \) the time step,  \( \Delta V \) the deviations of expectation values of the potential energy from its exact value  \( 0.025000 \) and \( \sigma \) the standard deviation from long simulations.}

\vskip 0.2cm \centering
\begin{tabular} {c c c c}
\hline \hline

N & $\Delta t$ & $\Delta V / 10^{-6} $ & $\sigma / 10^{-6}$ \\ [0.5ex]
\hline
  $10^4$ &   0.3    &   $160 $ & $540$ \\
  $10^4$ &   1      &   $ 60 $ & $530$ \\
  $10^4$ &   3      &   $ 40 $ & $470$ \\
$3 \times 10^4$ & 1 &   $ 30 $ & $320$ \\
\hline
\end{tabular}

\end{table}

\begin{figure}[ht]
\label{IMC}
\includegraphics[trim=2cm 0cm 2cm 1cm, clip=true,scale=0.45]{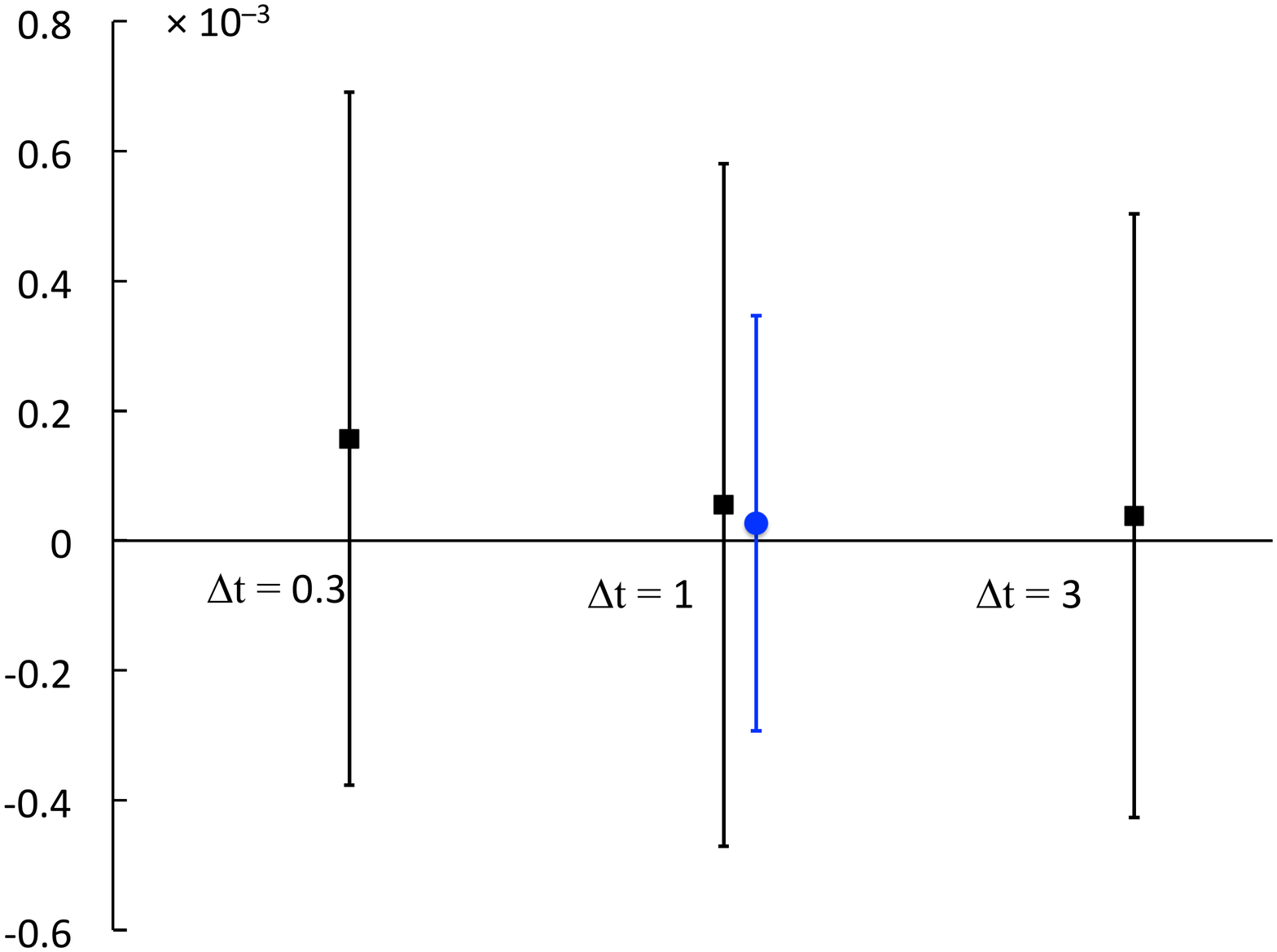}
\caption{Incoherent propagation in MC grid of the ODHO ground state with Trotter kernel.  Deviations of expectation values of the potential energy from its exact value 0.025 (dots) and standard deviations (bars) shown (in au \( \times 10^{-6} \)) from long simulations, with time steps \( 0.3 \), \( 1 \) and \( 3 \), and grid sizes \( 10^4 \) (black fullsquare) and \( 3 \times 10^4 \) (blue fullcircle). }
\end{figure}

It is worth noting that in a simulation, as described above, we have the ground state wave function at each step both in the walker distribution \( g(x) = \psi(a) \) and evaluated from propagation as \( \psi(b) \).  Though the latter is guiding the evolution of the former through MMC, \( g(x) \) can be kept stable by settings of the MMC parameters, whereas the stability of the evaluated amplitude \( \psi (b) \) depends primarily on the propagation parameters: grid size and time step length.  As a test case we present evaluation of the potential energy from Eq.~(\ref{ExptV}), which depends on both distributions.

To maximize variance (standard deviation) in this test, we use fully random and non overlapping grids \( {\bf g}_a \) and \( {\bf g}_b \) from exact gaussian distribution to assess the statistical performance of the Trotter kernel for evaluation of the ground state energetics of ODHO.  The obtained data from incoherent propagation is shown in Table 1 and Fig.~3.

We find that accuracy of the achieved ground state energetics (\( \Delta V \)) and distribution depends on the grid size and the time step.  Note, that the "error bars" (\( \sigma \)) do not describe accuracy.  Grid size dependence is as expected: larger grid increases accuracy.  Time step dependence, however, is weak and longer step leads to higher accuracy.  Overall, this what one can expect from the Trotter kernel.

The "error bars" in Fig.~3 describe simulation length independent standard deviation \( \sigma \) arising from Monte Carlo sampling.  It can be used to estimate the statistical accuracy (precision) of evaluated expectation values in form of standard error of mean, \( {\rm SEM} = \sigma / \sqrt{N_{\rm MC}} \), where \( {N_{\rm MC}} \) is the number of uncorrelated Monte Carlo steps.  Usually, \( 2 \times {\rm SEM} \) limits ( \( 95 \% \) ) are assumed as a statistical error estimate.  In our long simulations here, we found the real accuracy to be clearly worse than the statistical accuracy, due to the systematic error from Trotter approximation and such small test grid sizes.

\section{Conclusions}

We have demonstrated the path integral approach to the time domain coherent quantum dynamics with numerical simulations of simple one dimensional test cases, relevant as quantum dot models.  Generally, we find the PI approach more laborious as compared to the conventional evaluation of the solution from the time dependent Schr\"odinger equation, as expected \cite{feynman,feynman2}.

With PI approach a regular periodic grid may give rise to diffraction patterns on the evaluated amplitude, while Monte Carlo grids are free from such artifact.  Also as usual, with Monte Carlo technique for path sampling, the PI approach becomes more attractive in case of complex geometry or increasing number of spatial dimensions.

The cases where the exact kernel is known are special.  There, the time step length is not limited, even in practice, which offers a huge advantage over the conventional simulation of single particle quantum dynamics.  On the other hand, the straightforward incorporation of many-body correlations presumes short time steps.  Therefore, the Trotter kernel, which becomes exact at the zero step length limit, becomes accurate enough with practical time step lengths.  However, shorter time steps require more dense grids, as discussed above.

With the incoherent real time dynamics we have demonstrated a novel approach for searching the stationary states and the ground state, in particular.  Monte Carlo sampling of the continuous space turns out to increase accuracy as compared to the use of a regular discrete grid.  The Monte Carlo version has further advantages, similar to the conventional "high accuracy" diffusion Monte Carlo method.  Here, we have carried out the first tests of the convergence and accuracy of the new method, which seems promising with its novel features.

\section*{Acknowledgements}

For computational resources we like to thank the Techila Technologies facilities at Tampere University of Technology, and also, the facilities of Material Sciences National Grid Infrastructure (Akaatti, Merope) and Finnish IT Center for Science (CSC).
The authors also want to thank Dr.~Kyl\"anp\"a\"a for his comments on the manuscript.


\section*{References}
\begin{thebibliography}{10}
\bibitem{feynman} R.P.~Feynman and A.R.~Hibbs, Quantum Mechanics and Path Integrals (McGraw-Hill, New York, 1965).
\bibitem{feynman2} R.P.~Feynman, Rev.~Mod.~Phys.~{\bf 20}, 367 (1948).
\bibitem{kleinert} I.H.Duru and H.Kleinert, Phys.~Lett.~{\bf 84B}, 185 (1979) and H.Kleinert, Path Integrals in Quantum Mechanics, Statistics, Polymer Physics, and Financial Markets. World Scientific Publishing Co.~Pte.~Ltd.~Singapore (2004). The 5th edition.
\bibitem{schulman} L.S.Schulman, Techniques and Applications of Path Integration (Wiley, New York, 1981).
\bibitem{ceperley} D.M.~Ceperley, Rev.~Mod.~Phys.~{\bf 67}, 279 (1995).
\bibitem{kylanpaa} I.~Kyl\"anp\"a\"a, PhD Thesis (Tampere University of Technology 2011).
\bibitem{kylanpaaP} I.~Kyl\"anp\"a\"a and T.T.~Rantala, J.~Chem.~Phys. {\bf 133}, 044312 (2010), I.~Kyl\"anp\"a\"a and T.T.~Rantala, J.~Chem.~Phys.~{\bf 135}, 104310(2011) and I.~Kyl\"anp\"a\"a and T.T.~Rantala, Phys.~Rev.~A {\bf 80}, 024504(2009).
\bibitem{militzer} Militzer and D.M.~Ceperley, Phys.~Rev.~B {\bf 63}, 066404 (2001).
\bibitem{weiss} S.~Weiss and R.~Egger, Phys.~Rev.~B {\bf 72}, 245301 (2005).
\bibitem{gull} E.~Gull {\it et al.}, Rev.~Mod.~Phys.~{\bf 83}, 349 (2011).
\bibitem{makri1} N.~Makri, Comp.~Phys.~Comm.~{\bf 63}, 389--414.
\bibitem{makri2} N.~Makri, Chem.~Phys.~Lett.~{\bf 193}, 435 (1992).
\bibitem{filinov} V.S. Filinov, Nucl.~Phys.~B~{\bf 271}, 717--725 (1986).
\bibitem{wang} H.~Wang {\it et al.}, J.~Chem.~Phys.~{\bf 115}, 6317(2001).
\bibitem{makri3} N.~Makri, Ann.~Rev.~Phys.~Chem.~{\bf 50}, 167--191 (1999) and V.~Jadhao and N.~Makri, J.~Chem.~Phys.~{\bf 132}, 104110 (2010).
\bibitem{marchioro} T.L.~Marchioro and T.L.~Beck, J.~Chem.~Phys.~{\bf 96}, 2966 (1992).
\bibitem{makri4} N.~Makri, Comp.~Phys.~Comm.~{\bf 63}, 389--414 (1991) and N.~Makri, J.~Math.~Phys.~{\bf 36}, 2430--56 (1995).
\bibitem{makri5}  R.~Lambert and N.~Makri, J.~Chem.~Phys.~{\bf 137} 22A552 and 22A553 (2012).
\bibitem{makarov} D.E.~Makarov and N.~Makri, Chem.~Phys.~Lett.~{\bf 221}, 482 (1994).
\bibitem{ks} A.~Kolmogorov, G.Ist.Ital.Attuari {\bf 4}, 83 (1933).
\bibitem{suzuki} M.~Suzuki, Phys.~Lett.~A {\bf 201}, 425--428 (1995).
\bibitem{metropolis} N.~Metropolis, A.W.~Rosenbluth, M.N.~Rosenbluth, H.~Teller and E.~Teller, J.~Chem.~Phys.~{\bf 21}, 1087 (1953).
\bibitem{atkins} P.~Atkins and R.~Friedman, Molecular Quantum Mechanics (Oxford University Press Inc., New York, 2005). The 4th edition.
\bibitem{schulten} K.~Schulten, "Notes on Quantum Mechanics" (University of Illinois at Urbana–Champaign, 2000).



\end{thebibliography}
\end{document}